\documentclass[aps,prd,reprint,superscriptaddress,nofootinbib,showpacs,twoside,notitlepage]{revtex4-1}     
\usepackage{graphicx}
\usepackage{euscript,amssymb}
\usepackage{amsfonts}
\usepackage{amssymb}
\usepackage{amsmath}
\usepackage{color}

\newcommand{\be}{\begin{equation}}
  \newcommand{\ee}{\end{equation}}
\newcommand{\ben}{\begin{eqnarray*}}
  \newcommand{\een}{\end{eqnarray*}}
\newcommand{\bea}{\begin{eqnarray}}
  \newcommand{\eea}{\end{eqnarray}}
\newcommand{\bdm}{\begin{displaymath}}
  \newcommand{\edm}{\end{displaymath}}
\newcommand{\ba}{\begin{align}}
  \newcommand{\ea}{\end{align}}
\newcommand{\lb}{\label}

\newcommand{\I}{\text{i}}
\newcommand{\D}{\text{d}}

\newcommand{\dd}{\text{d}}
\newcommand{\dif}[3]{\frac{\partial^{#3} #1}{\partial #2^{#3}}}
\newcommand{\Ai}{\textnormal{Ai}}
\newcommand{\Bi}{\textnormal{Bi}}
\newcommand{\sgn}{\textnormal{sgn}}

\begin{document}

\title{Classical and quantum cosmology of Born-Infeld type models}

\author{Alexander Kamenshchik}

\email{alexander.kamenshchik@bo.infn.it}

\affiliation{Dipartimento di Fisica e Astronomia, Universit\`a di
  Bologna and INFN, Via 
Irnerio~46, 40126 Bologna, Italy}

\affiliation{L.D. Landau Institute for Theoretical Physics of the Russian
Academy of Sciences, Kosygin str. 2, 119334 Moscow, Russia}

\author{Claus Kiefer}

\email{kiefer@thp.uni-koeln.de}

\author{Nick Kwidzinski}

\email{nk@thp.uni-koeln.de}

\affiliation{Institut f\"ur Theoretische Physik, Universit\"{a}t zu
K\"{o}ln, Z\"{u}lpicher Stra\ss e 77, 50937 K\"{o}ln, Germany}

\date{\today}

\begin{abstract}
We discuss Born-Infeld type fields (tachyon fields) in classical and quantum cosmology. 
We first partly review and partly extend the discussion of the classical solutions
and focus in particular on the occurrence of singularities. 
For quantization, we employ geometrodynamics.
In the case of constant potential, we discuss both Wheeler-DeWitt quantization
and reduced quantization. We are able to give various solutions and discuss their
asymptotics. For the case of general potential, we transform the Wheeler-DeWitt
equation to a form where it leads to a difference equation.
Such a difference equation was previously found in the quantization of black holes.
We give explicit results for the cases of constant potential and inverse squared potential
and point out special features possessed by solutions of the difference equation.
\end{abstract}


\maketitle


\section{Introduction}\label{Intro}

The recent discovery  of cosmic acceleration \cite{cosmic} and 
the searches for Dark Energy, which can be responsible for such a phenomenon 
\cite{dark}, have stimulated studies of different cosmological models,
some of them including exotic types of fluids and fields. Among them are
the so called tachyon  
cosmological models \cite{Gibbons,Frolov, Fein,Padm,GKMP,Quiros}, which arise as
a by-product of string theory \cite{Sen}. The
energy-momentum tensor of the tachyon field  has a negative pressure
component which can be used for the description of the cosmic
acceleration. In spite of the somewhat misleading name,
these tachyon fields represent a development of the old idea by Born
and Infeld \cite{B-I} 
that the kinetic term of a field Lagrangian is not necessarily a
(quadratic) polynomial, but can contain a square root of 
fields and their derivatives.    

In the framework of modern cosmology, even more general Lagrangians
are employed, including some with arbitrary functions 
of the kinetic terms \cite{k}; these 
models are known as $k$-essence models. From our point of
view, however, the Born-Infeld 
type fields look more natural, because square-root Lagrangians arise 
in various parts of modern theoretical physics.
 The classical dynamics of 
tachyon Dark-Energy models is rich and can not only describe
cosmic acceleration, but can also lead to new
types of future singularities which are of interest in themselves \cite{GKMP, KGKGP, Kam-CQG}.   

The quantization of Born-Infeld type of models presents, however,
some challenge. Let us address the 
most popular quantization method for cosmological 
models, which is the construction of the wave function of the Universe
satisfying the Wheeler-DeWitt equation \cite{DeWitt,OUP}.
The main problem which one encounters there when applying this
framework to tachyonic models is the appearance of the momentum operators
under the square root \cite{KM12}. 
When we represent these operators as partial derivatives
of the field, we obtain non-local
differential operators, and one has to invoke sophisticated methods
for treating them.
 One possible method is to use the
analogy with the quantum mechanics of black holes and thin shells
developed in \cite{BKKT88,Berezin97,Hajicek92,BNT05}.   
 The corresponding Lagrangian contains  
the time derivatives of the observables under the square root, but 
the Hamiltonian depends on the momentum by a hyperbolic cosine. 
After quantization, this leads to a difference equation for the
wave function, which displays interesting features. 
 In the case of cosmology, the possible transition  
to difference equations does not arise automatically, but
can be achieved by means of an appropriate canonical
transformation. We shall discuss such equations in our paper.    

Besides the tachyonic field, there are also other
Born-Infeld type fields called pseudotachyons  
\cite{GKMP} and quasitachyons \cite{KGKGP}. Such fields arise in a
natural way in some cosmological models. In the model considered in
\cite{GKMP}, a particular potential containing the square root of a 
trigonometrical function was chosen. The dynamical evolution of the model can
bring the universe to a point where the expressions inside of 
the two square roots, in the potential  
and in the kinetic term, change sign simultaneously. Thus, to provide
a smooth cosmological evolution one is forced to change the Lagrangian
of the Born-Infeld type tachyon field, transforming  
it into the pseudotachyon field. 
After having crossed this point, the universe evolves towards a future 
cosmological singularity called Big Brake. This singularity is
characterized by a finite value of the cosmological radius of the
universe, by a vanishing Hubble parameter, and by an infinite
value of the cosmic deceleration. Singularities of such kind are
rather soft  
and can be passed through \cite{passing}; the details of the passage
of the Big Brake singularity in the model \cite{GKMP} are described
in \cite{passing1}.  In  \cite{KGK}, 
it was also noticed that the presence of dust matter in these
Big-Brake models can create additional difficulties.
In \cite{KGKGP}, it was shown that  
these difficulties can be overcome by means of another
Born-Infeld type field -  the
quasitachyon, which will be briefly mentioned in the next section.   
Some global aspects of quantum cosmology 
similar to the ones here were recently investigated in 
\cite{DKS,KKS,Paulo,KM12,Kam-CQG,Mariam,Bini16}. These concern, in particular,
the fate of classical singularities.

The structure of the paper is as follows. In Sec.~II, we present the
models for tachyonic and other Born-Infeld type fields and discuss 
their behavior in classical
cosmology. In Sec.~III, we address the quantum cosmology of these
fields in the Wheeler-DeWitt framework.
Sec.~IV is devoted to the alternative approach of reduced quantization.
In Sec.~V, we rewrite the Wheeler--DeWitt equation in the form of a
difference equation. We discuss various asymptotic forms and
show ways towards its solution.
The last section contains our conclusion.


\section{Tachyonic and other Born-Infeld type fields in classical
  cosmology}\label{classical} 

We shall work with flat Friedmann models given by the metric 
\begin{equation}
\D s^2=N^2(t)\D t^2-a^2(t)\D l^2,
\label{Fried}
\end{equation}
where $N(t)$ is the lapse function and $a(t)$ is the scale factor; we
choose $a$ to have the dimension of a length. 

The Lagrangian density for the spatially homogeneous tachyon field $T$ is 
\begin{equation}
L_T = -V(T)\sqrt{1-\frac{\dot{T}^2}{N^2}}, 
\label{Lagr}
\end{equation}
where $V(T)$ is the tachyon potential and the dot means time
derivative. The tachyon $T$ has the dimension of a length (and thus is
a geometric 
quantity), and $V$ has the dimension of a mass (energy) density; we set $c=1$. 

The energy density of the tachyon field is 
\begin{equation}
\rho = \frac{V(T)}{\sqrt{1-\frac{\dot{T}^2}{N^2}}},
\label{dens}
\end{equation}
while the pressure is
\begin{equation}
p=-V(T)\sqrt{1-\frac{\dot{T}^2}{N^2}}.
\label{pressure}
\end{equation}
We choose $V(T)\geq 0$ to have non-negative energy densities. 
We note that $p=-V^2(T)/\rho$. 

The total (minisuperspace) action is given by
\be
S={\mathcal V}_0\int \D t \left(-\frac{3a\dot{a}^2}{\kappa^2
    N}-Na^3V(T)\sqrt{1-\frac{\dot{T}^2}{N^2}}\right),
\ee
where $\kappa^2\equiv 8\pi G$, $G$ is the gravitational (Newton)
constant. The volume of three-space is ${\mathcal V}_0a^3$, where 
${\mathcal V}_0$ is a pure number that is set equal to one below. 

Choosing $N=1$, the total Lagrangian then reads
\begin{equation}
L = -a^3V(T)\sqrt{1-\dot{T}^2}-\frac{3a\dot{a}^2}{\kappa^2}.
\label{Lagr1}
\end{equation}

We have $\vert\dot{T}\vert\leq 1$ for the square root to stay real.
 From \eqref{Lagr1}, we get the equations of motion
\be
\lb{ddota}
\ddot{a}+\frac{\dot{a}^2}{2a}-\frac{a\kappa^2V}{2}\sqrt{1-\dot{T}^2}=0,
\ee
and
\be
\lb{ddotT}
\frac{\ddot{T}}{1-\dot{T}^2}+3H\dot{T}+\frac{V'(T)}{V(T)}=0,
\ee
where $H= \dot{a}/a$ is the Hubble parameter, and a prime denotes
a derivative with respect to the tachyon $T$. 

The canonical momenta read
\be
\label{momenta}
p_a=-\frac{6\dot{a}a}{\kappa^2},\quad
p_T=\frac{a^3V\dot{T}}{\sqrt{1-\dot{T}^2}}.
\ee

We note that both $p_a$ and $p_T$ have dimension of a mass. The usual
Legendre transform then yields the Hamiltonian

\be
\lb{H-constraint}
{\mathcal H}=-\frac{\kappa^2}{12}\frac{p_a^2}{a}+\sqrt{p_T^2+a^6V^2},
\ee
which is, in fact, a constraint, ${\mathcal H}=0$.

If expressed in terms of the velocities, this Hamiltonian constraint
gives the Friedmann equation
\be
\lb{Friedmann}
H^2=\frac{\dot{a}^2}{a^2}=\frac{\kappa^2}{3}\rho,
\ee
with $\rho$ given by \eqref{dens}. Using \eqref{dens} and \eqref{Friedmann} in
\eqref{ddota}, we get
\be
\ddot{a}=-\frac{\kappa^2aV}{2}\frac{3\dot{T}^2-2}{3\sqrt{1-\dot{T}^2}}.
\ee
With \eqref{dens} and \eqref{pressure}, this equation can be written
in the standard form
\be
\ddot{a}=-\frac{\kappa^2}{2}a(\rho+3p).
\ee

Let us first consider the special case of constant potential, 
$V(T)=V_0=$ constant. In this case, $T$ is a cyclic variable and $p_T$ thus a constant.
It was noticed in \cite{Frolov} that the corresponding cosmological
model is equivalent to a cosmological model with a Chaplygin gas \cite{Chap}, 
which has the equation of state 
\begin{equation}
p = -\frac{V_0^2}{\rho}.
\label{Chap}
\end{equation}

 From \eqref{ddotT} and \eqref{Friedmann}, one can then find the following
 equation for $\dot{T}^2$:
\be
\lb{dotT2}
\dot{T}^2=\frac{1}{1+\left(\frac{a}{a_*}\right)^6},
\ee
where $a_*$ is an integration constant,
\bdm
a_*:= \left(\frac{p_T}{V_0}\right)^\frac13.
\edm
We see that $\dot{T}^2$
vanishes for $a\to\infty$ and becomes equal to one for $a\to 0$ (Big
Bang). Using again \eqref{Friedmann}, \eqref{dotT2} can be integrated
to yield the curve in configuration space; one finds
\be T(a)=2\kappa^{-1}\sqrt{\frac{1}{3V}}\left(\frac{a}{a_*}\right)^{3/2}
{}_2\mathrm{F}_1\left(\frac14,\frac34;\frac54;-\left(\frac{a}{a_*}\right)^6
\right),
\label{config_traj}
\ee
where ${}_2\mathrm{F}_1$ denotes a hypergeometric function, see e.g. \cite{AS64}, Chap.~15.

\begin{figure}[h!]
	\centering
	\includegraphics[width=6.0cm]{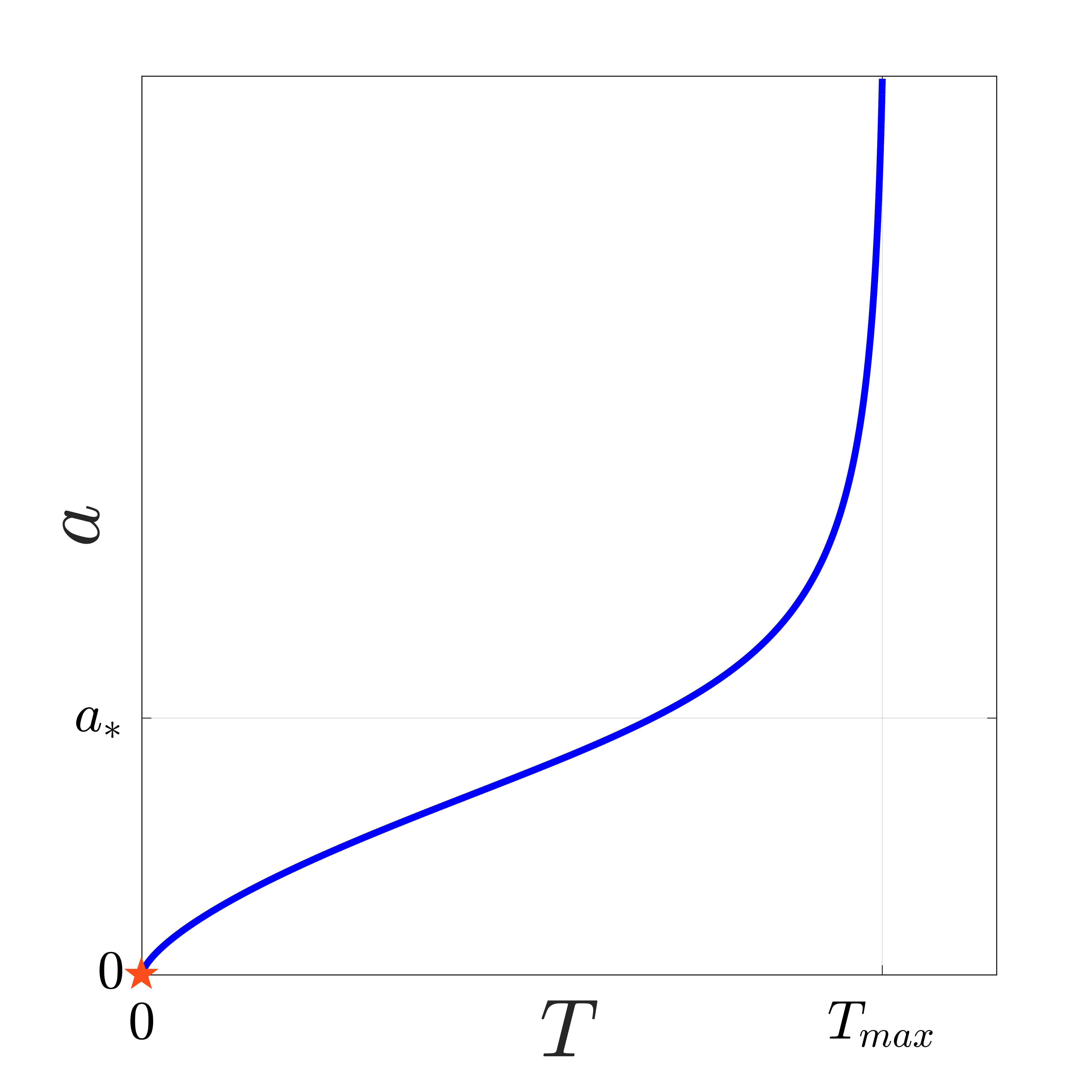}
	\caption{Plot of the configuration space trajectory (\ref{config_traj}) for the tachyon model with constant potential.}
\label{a_vs_T_Vconst}
\end{figure}

The asymptotic solution for large $a$ reads
\be
T(a)=\frac{1}{\kappa\sqrt{3V_0}}\left(\frac{a_*}{a}\right)^3+\mathrm{constant}.
\ee
Figure \ref{a_vs_T_Vconst} displays the configuration space trajectory \eqref{config_traj}.

 From \eqref{dens} and \eqref{pressure}, we get the following
 expressions for density and pressure:
\be
\label{rho_p}
\rho(a)=V_0\sqrt{1+\left(\frac{a_*}{a}\right)^6}, \quad
p(a)=-\frac{V_0^2}{\rho(a)}.
\ee

For $a\to 0$, both expressions diverge (Big Bang), while for
large $a$, they become constants with $p\approx -\rho$, thus mimicking
Dark Energy. We mention that for small $a$ the equation of state resembles the one
for dust. This model thus encodes a transition from a matter to a vacuum dominated
state, which is a feature observed in the real Universe.

Plugging the expression \eqref{rho_p} into the Friedman equation \eqref{Friedmann} yields 
\be
	\dot{a}^2=\frac{\kappa ^2 V_0}{3}\sqrt{a^4 +\frac{a_*^6}{a^2}},
\ee
which is solved by

\begin{align}
\sqrt{\frac{\kappa ^2 V_0}{3}}(t-t_0)=&\frac{2a^3}{3a_*^6}
\left( a^4+ \frac{a_*^6}{a^2}\right)^{\frac{3}{4}} {}_2F_1 \left(1,1;\frac{5}{4};-\left(\frac{a}{a_*}\right)^6 \right) \nonumber \\
& + \text{constant}.
\label{a_of_t_V}
\end{align}

\begin{figure}[h!]
	\centering
	\includegraphics[width=6.0cm]{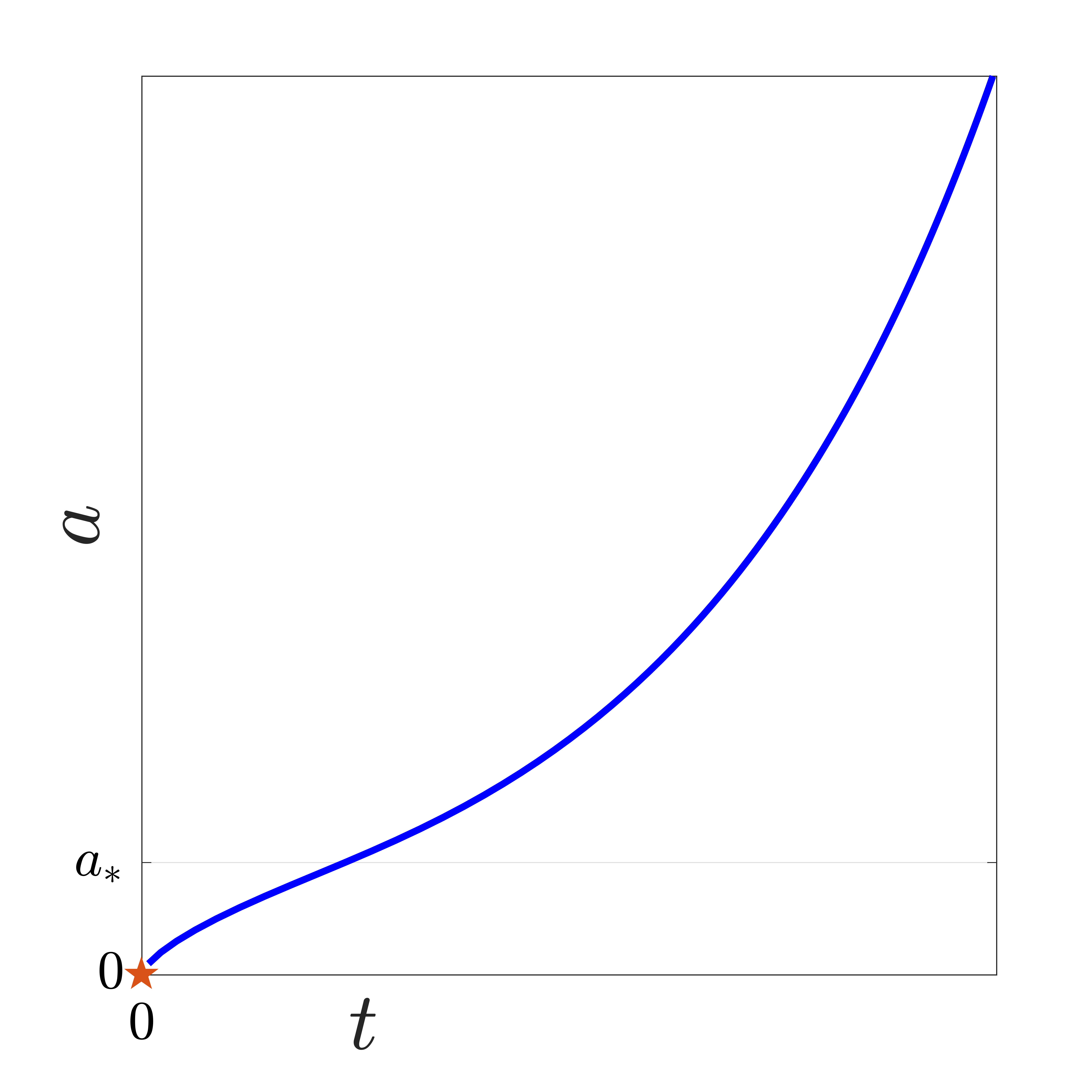}
	\caption{Plot of the general solution $a(t)$ for the tachyon model with constant potential.}
\label{a_Vconstant.pdf}
\end{figure}

The plot of this trajectory is displayed in Fig.~\ref{a_Vconstant.pdf}.

\begin{figure}[h!]
	\includegraphics[width=8.6cm]{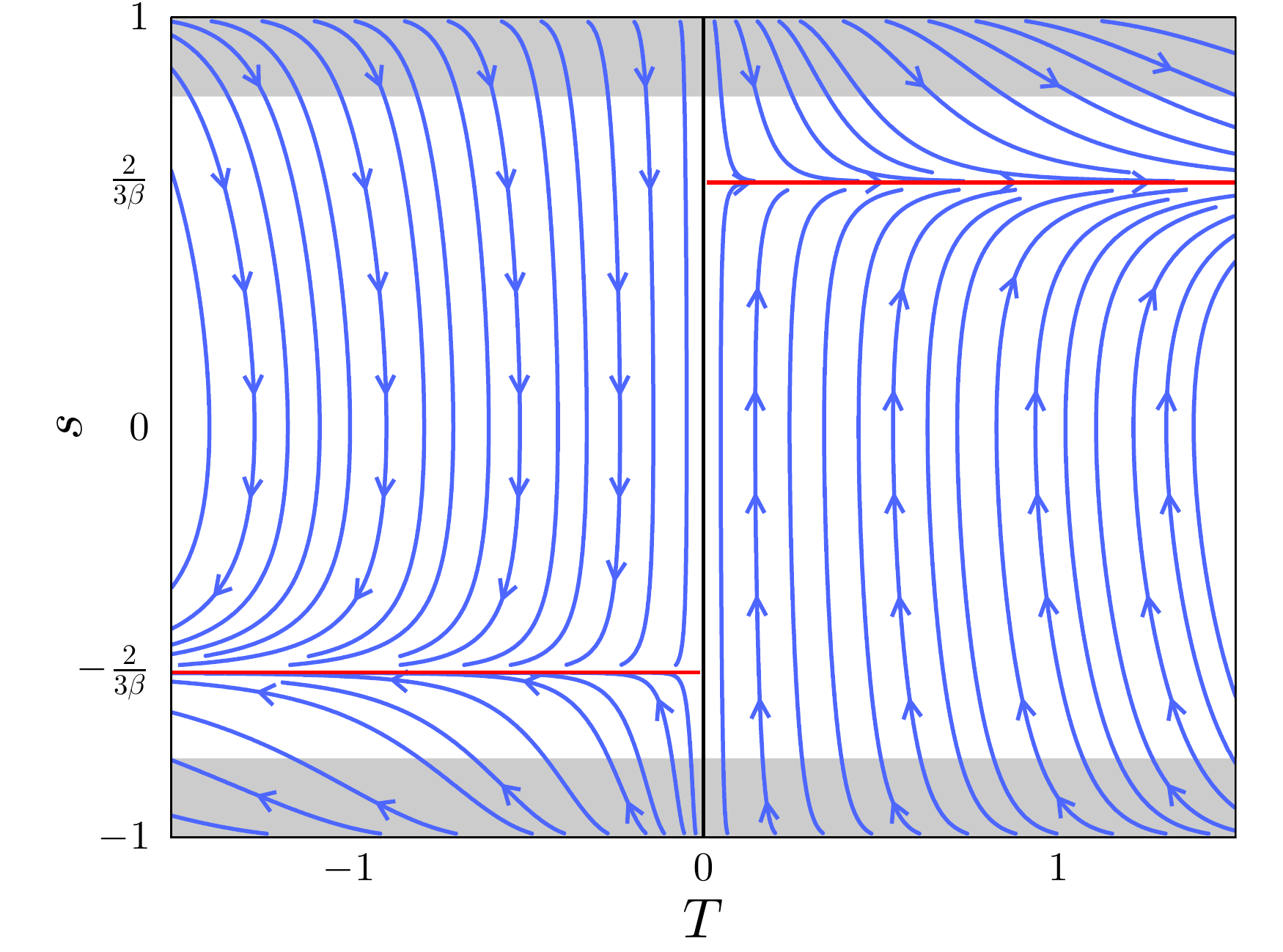}
	\caption{Flow diagram of the tachyon model with the inverse square potential. Inside the grey shaded region,
	the universe undergoes a decelerated expansion, while it accelerates outside this region.
	The parameter $\beta$ is given by $\beta = \frac{1}{9}\sqrt{2+\sqrt{9 \kappa^4 V_1^2 + 4}}$;
	  $T$ is in arbitrary units of time.}
\label{flow_V}
\end{figure}

Also of interest is the model of an inverse square potential, $V={V_1}/{T^2}$; see, for example
\cite{Quiros,Copeland}. 
This model exhibits a particular solution with constant $\dot{T}$, describing a universe that undergoes 
a power-law-inflation \cite{Fein,Padm}.  This can be seen as follows.
From \eqref{dens}, (\ref{ddotT}), and (\ref{Friedmann}), one obtains the dynamical system
\begin{equation}
	\frac{\dd}{\dd t}\left(\begin{array}{c}
	T \\
	s
	\end{array} \right)	= 
	\left(\begin{array}{c}
	s \\
	-\left(1-s^2\right)\frac{V'(T)}{V(T)} - \kappa s\sqrt{3V(T)}\left(1-s^2\right)^{\frac{3}{4}}  
	\end{array} \right),
\end{equation}
where $s:=\dot{T}$. The corresponding flow diagram is depicted in Fig.~\ref{flow_V}.
One recognizes from the diagram that the particular solution with constant $\dot{T}$ serves as an attractor.
For other tachyon potentials, see for example \cite{Quiros}. 

We shall now address another type of Born-Infeld-type field
called ``pseudotachyon'' which, as was explained in the Introduction,
naturally arises in cosmological models \cite{GKMP}; it is less habitual than 
the tachyon model, but displays some interesting theoretical features.
Its Lagrangian density reads
\begin{equation}
 L_p = W(T) \sqrt{\dot{T}^2-1}.
 \label{Lagr2}
\end{equation}
The Friedmann equation and the Klein-Gordon type equation are given by
\begin{equation}
	H^2=\frac{\dot{a}^2}{a^2}
	=\frac{\kappa ^2}{3}\frac{W(T)}{\sqrt{\dot{T}^2-1}}
	\label{Friedmann_W}
\end{equation}
and
\begin{equation}
\frac{\ddot{T}}{1-\dot{T}^2}+3H\dot{T}+\frac{W'(T)}{W(T)}=0,
\label{KG_W}
\end{equation}
respectively.
Energy density and pressure of the pseudotachyon read
\begin{equation}
\rho = \frac{W(T)}{\sqrt{\dot{T}^2-1}}
\quad \text{and} \quad 
p=W(T)\sqrt{\dot{T}^2-1}.
\label{dens2}
\end{equation}

\begin{figure}[h!]
	\centering
	\includegraphics[width=6.0cm]{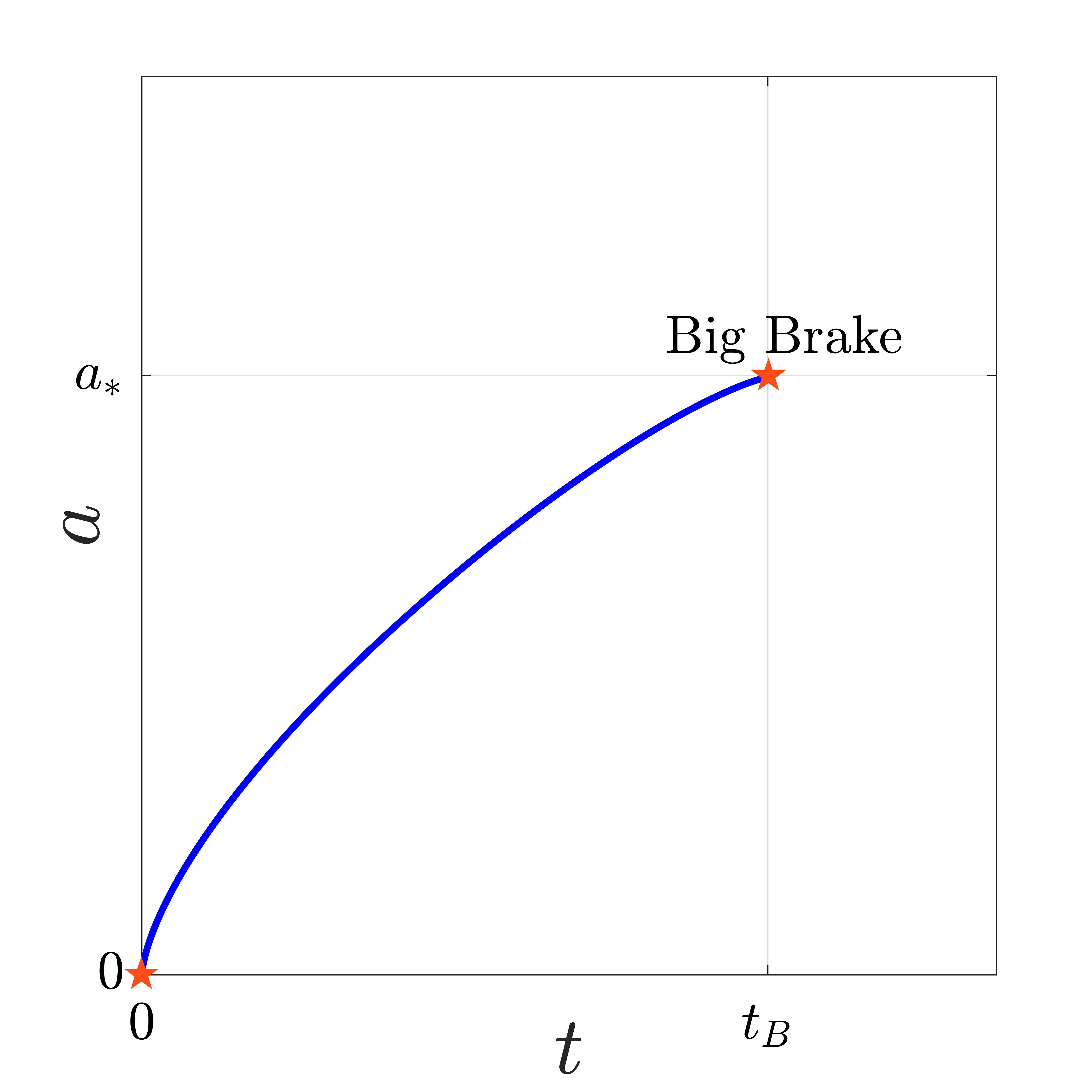}
	\caption{Plot of the general solution $a(t)$ for the pseudotachyon model with constant potential.}
\label{a_vs_t_Wconst}
\end{figure}

\begin{figure}[h!]
	\centering
	\includegraphics[width=6.0cm]{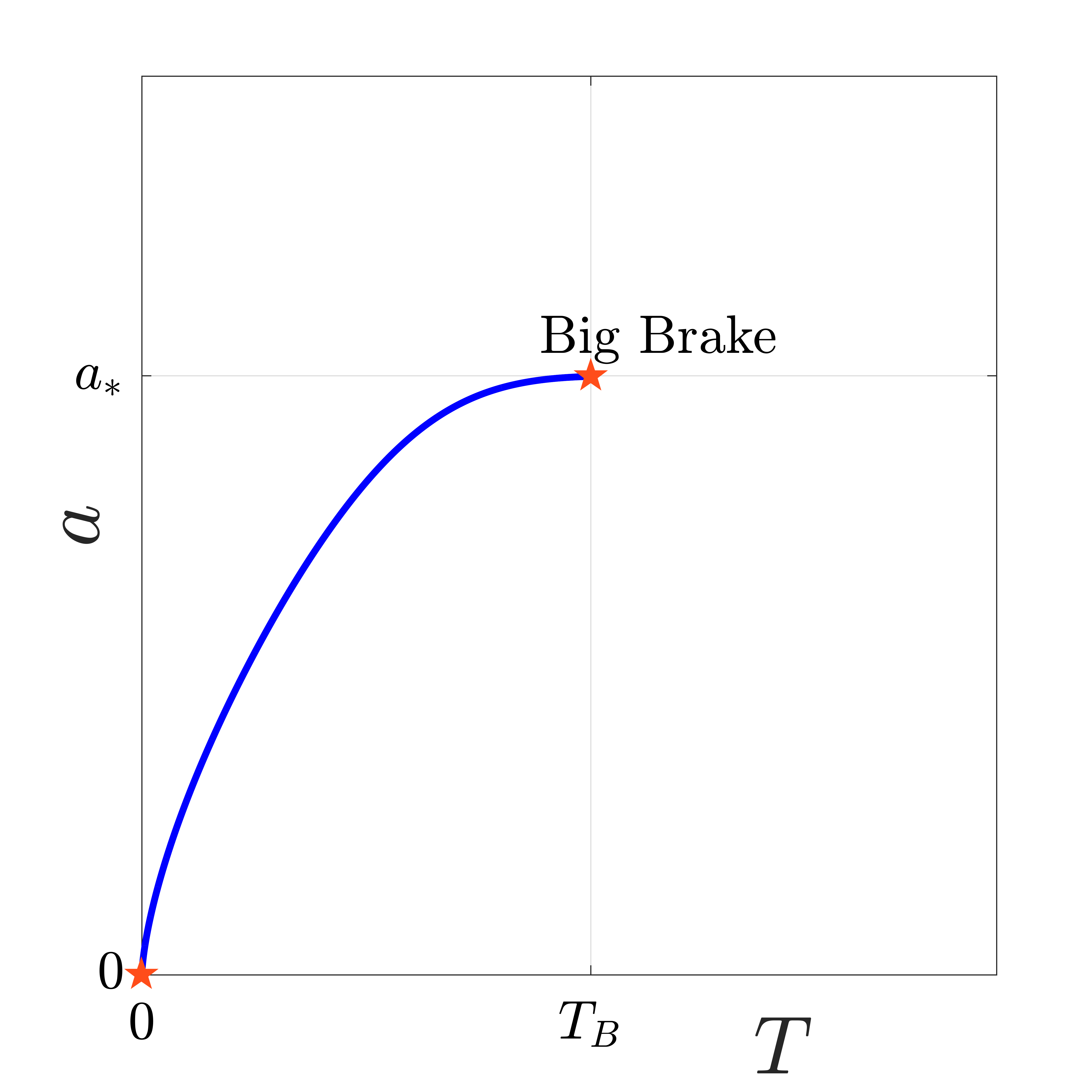}
	\caption{Plot of configuration space trajectory for the pseudotachyon model with constant potential.}
\label{a_vs_T_Wconst}
\end{figure}

In this case, both the energy density and the pressure are positive.
When the potential is constant, $W(T)=W_0=$ constant, this model coincides with
a cosmological model containing 
an anti-Chaplygin gas \cite{GKMP,KKS}; this is a perfect fluid with the equation
of state 
\begin{equation}
p = \frac{W_0^2}{\rho}.
\label{anti-Chap}
\end{equation}
A universe with an anti-Chaplygin gas represents the simplest example
of a cosmological evolution with a Big Brake singularity. It is
interesting that the equation of state (\ref{anti-Chap}) arises in the
theory of wiggly strings \cite{wiggly}. 
In Fig.~\ref{a_vs_t_Wconst}, we depict the trajectory $a(t)$ and in Fig.~\ref{a_vs_T_Wconst} the trajectory in 
configuration space. We see explicitly the occurrence of the Big Brake singularity.

The Hamiltonian is readily obtained by a Legendre transform,
\be
\lb{H-constraint_W}
{\mathcal H}=-\frac{\kappa^2}{12}\frac{p_a^2}{a}+\sqrt{p_T^2-a^6W^2}.
\ee

\begin{figure}[h!]
	\includegraphics[width=8.6cm]{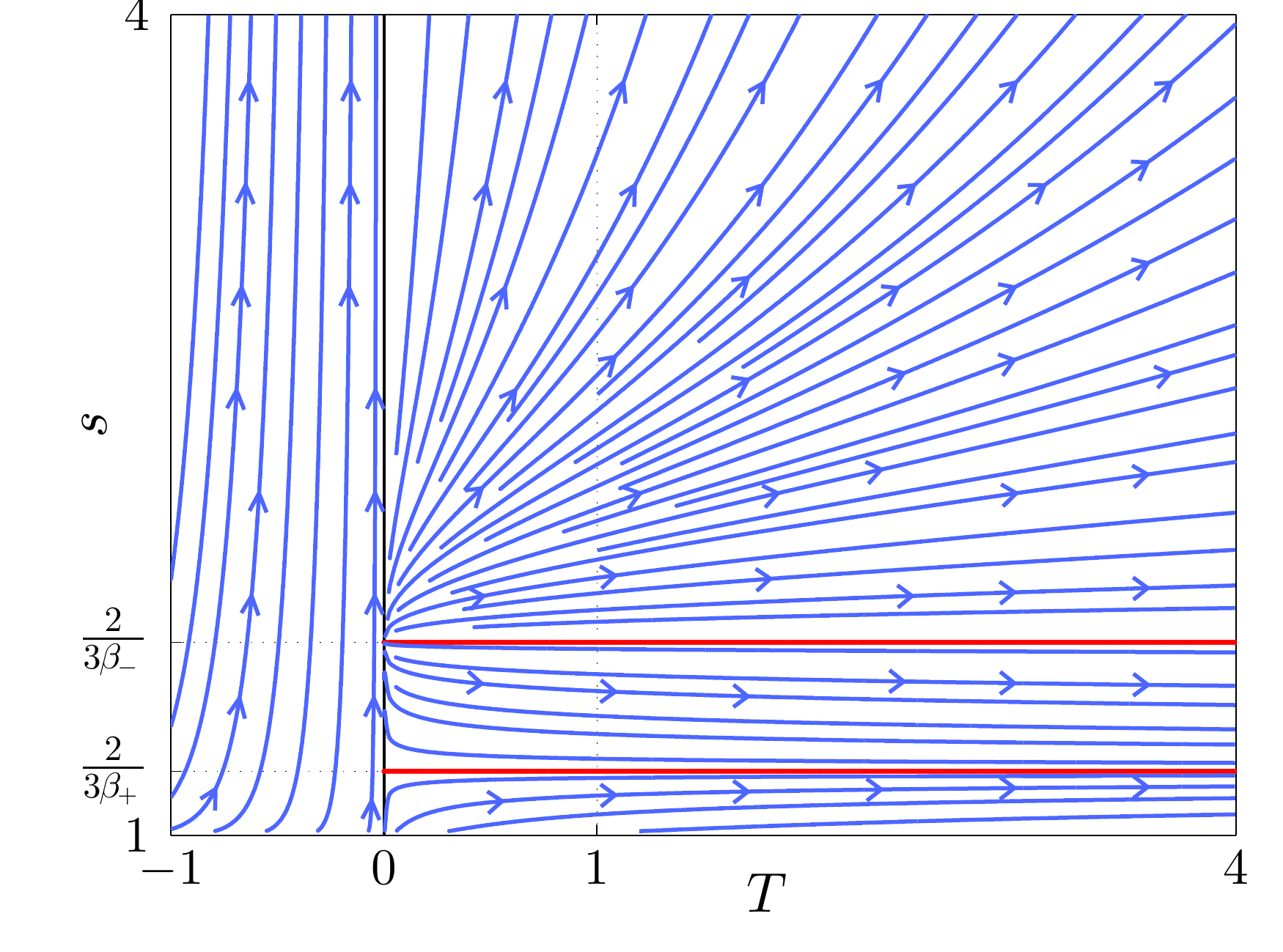}
	\caption{Flow diagram of the pseudotachyon model with the inverse square potential for the case $W_1<\frac{2}{9}$;
	   $T$ is in arbitrary units of time.}
\label{flow_W1}
\end{figure}

In the following, we consider the case of the inverse square potential, $W={W_1}/{T^2}$. 
This does not seem to have been discussed so far in this way. 
For simplicity, we set here ${\kappa^2}/{3}=1$. 
Similarly to the corresponding tachyon model we find solutions with constant $\dot{T}$. In this case, however,
we have two solutions $\dot{T}_\pm=\frac{2}{3\beta_{\pm}}$, where $\beta_{\pm}^2=
	\frac{1}{9}\left( 2 \pm \sqrt{4-81 W_1^2 } \right)$.
In order to get real solutions, we have to demand $W_1\leq\frac{2}{9}$. The scale factors
for these two solutions are given by
\begin{equation}
\label{apm}
	a_{\pm}(t) \propto t^{\frac{3\beta_{\pm} ^2}{2}}.
\end{equation}
In the limiting case $W_1=\frac{2}{9}$, the two solutions merge into one. 
Analogously to the tachyon case, one can express the dynamics in the form
\begin{equation}
\label{dynamical}
	\frac{\dd}{\dd t}\left(\begin{array}{c}
	T \\
	s
	\end{array} \right)	= 
	\left(\begin{array}{c}
	s \\
	\left(s^2-1\right)\frac{W'(T)}{W(T)} +
	 3 s\sqrt{W(T)}\left(s^2-1\right)^{\frac{3}{4}}  
	\end{array} \right).
\end{equation}
The flow chart for the case $W_1<\frac{2}{9}$ is shown in Fig.~\ref{flow_W1}. 
A closer inspection reveals that all solutions emerge from a Big Bang on the line determined by $s=1$ or
	the single point $(T=0,s=\frac{2}{3\beta_-})$. One explicitly sees that one of the particular solutions 
	($s=\frac{2}{3\beta_+}$)
	serves as an attractor, while the other one ($s<\frac{2}{3\beta_-}$)
	is repulsive. The solutions in the region $T>0$, $s<\frac{2}{3\beta_-}$ 
	are attracted to the particular solution with $s=\frac{2}{3\beta_+}$,
	cf. \eqref{apm}. All other solutions end in a Big Brake. 
	In the limiting case $W_1=\frac{2}{9}$, the two particular solutions merge into one metastable solution.
	For $W_1>\frac{2}{9}$ (not shown here), the particular solutions disappear, and all solutions end in a Big Brake. 
	
In the following, we shall explicitly show the presence of the Big Brake singularity for a typical case.
By defining $u:=\ln \left(\sqrt{\dot{T}^2-1} \right)$, one can find from (\ref{dynamical}),
\begin{equation}
	\frac{\dd u}{\dd T}=\frac{1}{T}\left[ 3\ \sgn \left( T \right)
    \sqrt{2W_1\cosh\left( u\right)}-2 \right].
	\label{KG_W2}
\end{equation}
Integration then yields a parametrization of $T$ in terms of its time derivative $\dot{T}$, 
\begin{equation}
	T=T(\dot{T}_r)\exp \left[- 
		\int\limits _{\ln \left( \sqrt{\dot{T}^2_r-1}\right)}
		^{\ln \left( \sqrt{\dot{T}^2-1} \right) }
		\frac{\dd u}{2-3\ \sgn (T)\sqrt{2 W_1 \cosh (u)} }
	\right],
	\label{paramtrization_T}
\end{equation} 
where $T(\dot{T}_r)$ is the value of $T$ at some reference value $\dot{T}_r$.
This parametrization can now be used to prove the existence of the singularities. 

To be specific, we consider the case $W_1<\frac{2}{9}$ and the solutions in the region 
where $T>0$ and $\dot{T} > \frac{2}{3\beta_-}$.
We shall now show by using suitable estimates
of the integral in (\ref{paramtrization_T}) that $T\rightarrow 0$ as $\dot{T}\rightarrow \frac{2}{3\beta_-}$ 
and that $T$ approaches a finite value $T_\infty$ as $\dot{T}\rightarrow \infty$;
here, it is convenient to use $\dot{T}_r=\infty$ as a reference value. 	
Expression (\ref{paramtrization_T}) then becomes
 \begin{equation}
	T=T_{\infty}\exp \left[ 
		\int\limits_{\infty}^{\ln \left( \sqrt{\dot{T}^2-1}\right) }\frac{\dd u}{3\sqrt{2 W_1 \cosh (u)}-2 }
	\right].
	\label{paramtrization_T2}
\end{equation}
If we divide the function inside the integral by the function
\begin{equation}
\frac{1}{3\sqrt{ W_1 e^u}-2 },
\label{aux_func}
\end{equation}
the resulting function approaches 1 as $u\rightarrow\infty$.
The integral of (\ref{aux_func}) over the same interval as in (\ref{paramtrization_T2}) 
is finite for $\dot{T} > \frac{2}{3\beta_-}$.  Consequently, 
we can use the limit comparison test to conclude that the expression (\ref{paramtrization_T2}) is well defined and therefore
$T$ approaches a finite value $T_\infty$ as $\dot{T}\rightarrow \infty$. 

If we do the re-substitution 
$u=\ln\left( \sqrt{s^2-1}\right)$, the integral in (\ref{paramtrization_T2}) assumes the form 
\begin{equation}
	\int_{\infty}^{\dot{T}}\frac{\dd s}{\left(s^2-1 \right)
	\left(3s\sqrt{\frac{W_1}{\sqrt{s^2-1}}}-2 \right)}.
\end{equation}
If we now choose $s\in [s_-,s_- +\varepsilon)$, where $s_-:=\frac{2}{3\beta_-}$ and $\varepsilon>0$,
we can estimate the integrand to be bigger than
\begin{equation}
\frac{1}{\left( (s_-+\varepsilon)^2-1\right)^\frac{3}{4}\left(3\sqrt{W_1}s-2(s_-^2-1)^\frac{1}{4} \right)}.	
\label{aux}
\end{equation}
By noting that $s_-$ is a zero of the denominator, we conclude that the integral of this expression
over the interval $[s_-,s_- +\varepsilon)$ blows up to $+\infty$. Therefore,
the integral in (\ref{paramtrization_T2}) goes to $-\infty$ as $\dot{T}\rightarrow \frac{2}{3\beta_-}$
and thus $T\rightarrow 0$.
By estimating that 
\begin{equation}
T_\infty = \int_{t_0}^{t_\infty}\dd t \ \dot{T}> t_\infty - t_0,
\end{equation} 
with $t_\infty$ corresponding to $T_\infty$ and $t_0$ corresponding to $T=0$,
we deduce that $T$ grows from $0$ to $T_\infty$ in a finite amount of time.
Later on, we show that this model possesses a constant of motion, see (\ref{const_of_motion}) below.
This relation can be written as
\begin{equation}
	a^3=\frac{C}{\frac{3W_1 \dot{T}}{T\sqrt{\dot{T}^2-1}}-2H},
\end{equation}
where $C$ is a positive constant.
 The considerations above now yield, on the one hand,
\begin{equation}
 a\rightarrow 0,\ \rho\rightarrow\infty \ \text{and} \ p\rightarrow \infty \quad \text{as} 
 \quad  \dot{T}\rightarrow \frac{2}{3\beta_-}
\end{equation}
 (Big Bang) and, on the other hand, 
 \begin{equation}
 a\rightarrow\tfrac{C}{3W_1}T_\infty,\ \rho\rightarrow 0 \ \text{and} \ p\rightarrow \infty 
 \quad   \text{as} \quad \dot{T}\rightarrow \infty
 \end{equation}
 (Big Brake).
 The limit $p\rightarrow \infty$ now implies that $\ddot{a}\rightarrow-\infty$, 
 and we finally conclude that the solutions start from a Big Bang and end in a Big Brake. 
 
\begin{figure}[h!]
	\centering
	\includegraphics[width=6.0cm]{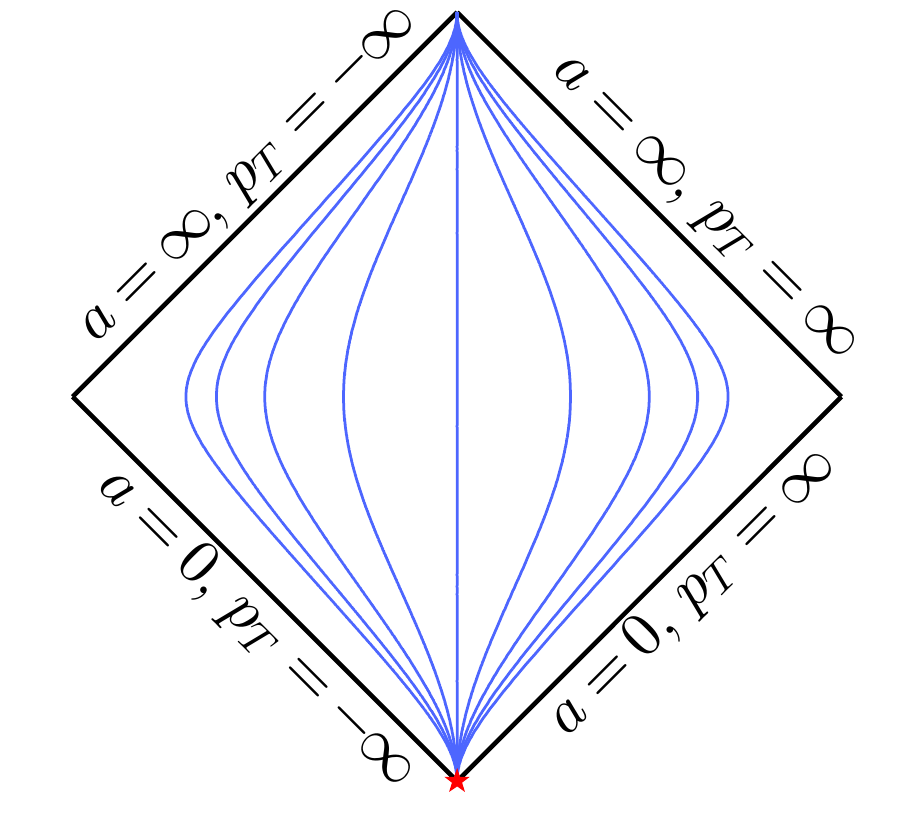}
	\caption{Phase space trajectories of the constant potential tachyon
		field models. All solutions (except the one in the middle) evolve out of a Big Bang singularity
		marked by the red star and end in the point $\left(a=\infty,p_T\right)$.}
\label{conformal_V}
\end{figure}

\begin{figure}[h!]
	\centering
	\includegraphics[width=6.0cm]{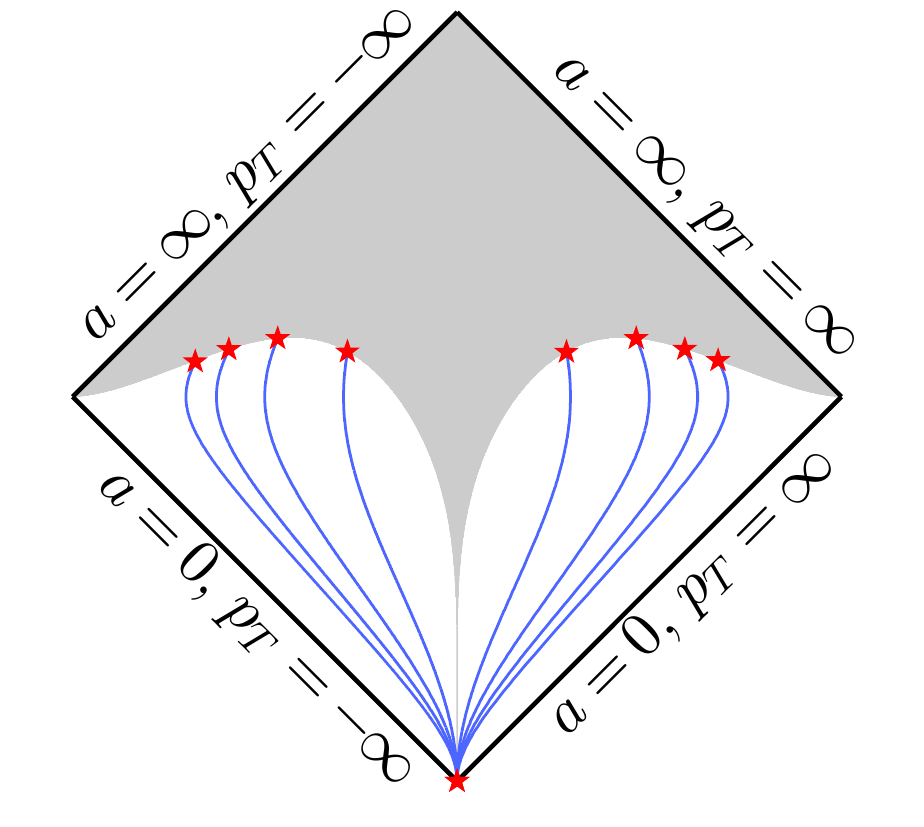}
	\caption{Phase space trajectories of the constant potential pseudotachyon
		field models. All solutions evolve out of a Big Bang singularity marked by the red star
		and end in a Big Brake singularity at the edge of the grey shaded region,
		where the Hamiltonian (\ref{H-constraint_W}) becomes ill defined. 
		In the case of a contracting universe, the trajectories go along the same lines,
		starting from the Big Brake towards the Big Crunch singularity.}
\label{conformal_W}
\end{figure}

 In the case of constant potential, one can employ conformal diagrams to illustrate the behavior
 of solutions. Diagrams of this kind have been used in quantum cosmology before; see, for example,
 \cite{Wiltshire}. Figure~\ref{conformal_V} shows the case of the tachyon, while Fig.~\ref{conformal_W}
 shows the case of the pseudotachyon field. 
The blue lines mark the trajectories
in the $(a,p_T)$- phase space corresponding to 
lines with constant $p_T$. Here, $V_0=W_0=\frac{1}{2}$, respectively.

In some particular cases, yet another Born-Infeld type field can
arise \cite{KGKGP}, with the Lagrangian density 
\begin{equation}
 L_q = U(T) \sqrt{\dot{T}^2+1}.
 \label{Lagr3}
\end{equation}
In this case, the energy density is negative, while the pressure is
positive. This field is called ``quasitachyon''. When $U(T)=U_0=$
constant, the
quasitachyon field behaves like a Chaplygin gas with negative energy
density and positive pressure. In this paper, however, we shall restrict attention to the tachyon 
and the pseudotachyon fields.  

We now turn to the quantum versions of the tachyon and pseudotachyon models.

\section{Quantum cosmology for Born-Infeld type fields} 

In spite of their apparently simple character, already the 
models with constant potentials are
rather complicated from the point of view of quantum cosmology. 
In the following, we shall discuss various approaches for their
quantization. 

 From \eqref{H-constraint}, we get the following Wheeler-DeWitt
 equation  for a universe filled with a tachyonic 
field \cite{KM12}, 
\begin{equation}
\left(\sqrt{p_T^2+a^6V^2}-\frac{p_a^2}{2a}\right)\Psi(T,a)=0,
\label{WDW0}
\end{equation}
where $\Psi(T,a)$ is the quantum state of the universe and $p_T,T,a$,
and $p_a$ are now operators; here and in the following we set
$\kappa^2=6$.  

Equations such as \eqref{WDW0} are plagued by the factor-ordering problem:
there is no unique way to transform the classical configuration and momentum
variables into operators \cite{OUP}. Here, we shall adopt a pragmatic attitude
and choose a simple factor ordering which facilitates the finding
of explicit solutions. 

Usually, one implements $a$ and $T$ as multiplication operators and
$p_a$ and $p_T$ as derivative operators.  
In view of the square root in \eqref{WDW0}, this is, however, a delicate issue.
But in the particular case of constant potential, we can use the
fact that the field 
$T$ does not enter explicitly into the Wheeler-DeWitt equation. Thus,
we can use a momentum representation for the tachyon and consider instead of  
the wave function $\Psi(T,a)$ the wave function $\Psi(p_T,a)$
(using, for simplicity, the same letter). 
In this case,  
the operator $p_T$ becomes multiplicative, and the Wheeler-DeWitt
equation acquires the form 
\begin{equation}
\left(\frac{\partial^2}{\partial
    a^2}+2a\sqrt{p_T^2+a^6V_0^2}\right)\Psi(p_T,a)=0.  
\label{WDW1}
\end{equation}
We can look for a solution of (\ref{WDW1}) in the form 
\begin{equation}
\Psi(p_T,a) = \psi(p_T,a)\chi(p_T),
\label{WDW2}
\end{equation}
where 
$\chi(p_T)$ denotes an arbitrary function of $p_T$. 
In this case, we arrive at
\begin{equation}
\left(\frac{\partial^2}{\partial
    a^2}+2a\sqrt{p_T^2+a^6V_0^2}\right)\psi(p_T,a)=0,  
\label{WDW3}
\end{equation}
where $p_T$ is a fixed parameter. The solutions of this equations do
not seem to belong to  
known special functions, but we can consider some limiting
cases. Namely, in the case 
when the cosmological radius is small, we have
\begin{equation}
\left(\frac{\partial^2}{\partial a^2}+2a|p_T|\right)\psi(p_T,a)=0. 
\label{WDW4}
\end{equation}
The solution of this equation is known; it can be expressed by means of
Airy functions: 
\begin{equation}
\psi(p_T,a)=c_1\mathrm{Ai}((-2|p_T|)^{1/3}a)+c_2\mathrm{Bi}((-2|p_T|)^{1/3}a).
\label{Airy} 
\end{equation}
However, because (\ref{WDW4}) is valid only in the limit $a
\rightarrow 0$, we need to take into account in the solution (\ref{Airy}) only the
leading terms and rewrite it as  
\begin{equation}
\psi(p_T,a)=d_1+d_2 a.
\label{Airy1}
\end{equation}
Because $a=0$ corresponds in the classical model to the Big Bang, the
question of singularity avoidance in quantum 
cosmology can be addressed. DeWitt has proposed the heuristic
criterion that the wave function should vanish at the point of the
classical singularity \cite{DeWitt}. This criterion was implemented in
the models discussed in \cite{KKS,Paulo,Mariam}. If we adopt this
criterion here, we have to demand that $\psi(p_T,0)=0$, that is, we
have to choose $d_1=0$. For this choice, then, the Big Bang
singularity would be avoided in the sense of DeWitt.

When $a$ is very large, we get from \eqref{WDW3} the following equation, 
\begin{equation}
\left(\frac{\partial^2}{\partial a^2}+2a^4V_0\right)\psi(p_T,a)=0. 
\label{WDW5}
\end{equation}
Its solution can be expressed in terms of Bessel functions,
\begin{equation}
\psi(p_T,a)=f_1\sqrt{a}J_{-1/6}\left(\frac{\sqrt{2V_0}a^3}{3}\right)
+f_2\sqrt{a}J_{1/6}\left(\frac{\sqrt{2V_0}a^3}{3}\right).
\label{Bessel}
\end{equation}
(Recall that $V_0\geq 0$.)
This is the quantum solution that corresponds to the asymptotic
de~Sitter phase of the classical solution, which is well known from
the solution of the Wheeler-DeWitt equation with a cosmological
constant (\cite{OUP}, Chap.~8). 

Keeping only the leading terms at $a
\rightarrow \infty$, this becomes
\begin{equation}
\psi(p_T,a)=g_1\exp\left(\I\frac{\sqrt{2V_0}a^3}{3}\right)+
g_2\exp\left(-\I\frac{\sqrt{2V_0}a^3}{3}\right).  
\label{Bessel1}
\end{equation}

Let us now consider the pseudotachyon field with constant potential. 
In this case, the Wheeler-DeWitt equation has the following form:
\begin{equation}
\left(\frac{\partial^2}{\partial
    a^2}+2a\sqrt{p_T^2-a^6W_0^2}\right)\Psi(p_T,a)=0.  
\label{pWDW1}
\end{equation}
With an ansatz of the form \eqref{WDW2}, we get
\begin{equation}
\left(\frac{\partial^2}{\partial
    a^2}+2a\sqrt{p_T^2-a^6W_0^2}\right)\psi(p_T,a)=0.  
\label{pWDW2}
\end{equation}
At small values of $a$, this equation coincides with (\ref{WDW4}) and
thus leads to the same solution in this limit. 

The value of the scale factor 
\begin{equation}
a_*:=\left(\frac{|p_T|}{W_0}\right)^{\frac{1}{3}}
\label{brake}
\end{equation}
corresponds to the Big Brake singularity. Let us consider the
Wheeler-DeWitt equation (\ref{pWDW2}) in  the neighborhood of this
point and write for this purpose
\begin{equation}
a=:a_*-\tilde{a}.
\label{brake1}
\end{equation}
We then have
 \begin{equation}
 \left(\frac{\partial^2}{\partial
     \tilde{a}^2}+2\sqrt{6}W_0(a_*)^{7/2}\sqrt{\tilde{a}}\right)\psi(p_T,\tilde{a})=0.  
 \label{brake2}
 \end{equation}
 Its solution is 
 \begin{eqnarray}
 &&\psi(p_T,a)=c_1\sqrt{\tilde{a}}
J_{-2/5}\left(\frac45\sqrt{2\sqrt{6}W_0(a_*)^{7/2}}(\tilde{a})^{5/4}\right)\nonumber
 \\ 
&&
+c_2\sqrt{\tilde{a}}
J_{2/5}\left(\frac45\sqrt{2\sqrt{6}W_0(a_*)^{7/2}}(\tilde{a})^{5/4}\right). 
 \label{brake3}
 \end{eqnarray}
For small values of $\tilde{a}$, it behaves as 
\begin{equation}
 \psi(p_T,a)=d_1+d_2\tilde{a}.
 \label{brake4}
 \end{equation}
The self-adjointness of the Hamiltonian operator in the Wheeler-DeWitt
equation is an open issue \cite{OUP}. But if we demand this property
to hold here, $\tilde{a}$ cannot be negative because otherwise
 the expressions under the square roots
 in (\ref{pWDW2}) and (\ref{brake2}) would become negative.
We thus have to impose the boundary condition
\begin{equation}
\psi(p_T,a) = 0\ \ {\rm at}\ \ \tilde{a} \leq 0.
\label{brake5}
\end{equation}
According to the DeWitt criterion, the Big Brake singularity is
then avoided, too. This is similar to the avoidance found in \cite{KKS,Paulo,Mariam}.

The question of the reality of the spectrum of the Hamiltonian in
quantum cosmology was considered already in 
\cite{Blyth-Isham}. There, another approach to the construction of the
wave function of the universe called reduced
quantization was discussed \cite{OUP}. In this approach, a time
parameter is chosen from the classical phase space variables, and a
non-zero Hamiltonian appears, which depends on this 
time parameter and the physical degrees of freedom in the reduced
phase space of the theory. Upon quantization, one arrives at a 
Schr\"odinger equation for the wave function of the universe,
depending on time and the physical degrees of freedom. In this case,
the Hamiltonian almost unavoidably contains square roots, even if the initial
Lagrangian does not contain them. This, together with other problems,
makes the reduced approach untractable in most cases \cite{OUP,Kuchar}. 

Later, the reduced approach was developed in
great detail in \cite{barvin}, and its relation with Dirac
quantization approach was analyzed. Its application to some rather
simple cosmological models was  
presented in the recent paper \cite{BarKam2014}.   
However, considering the Born-Infeld type models, we encounter a more
complicated problem, because here the square-root type Hamiltonians
are present in the Wheeler-DeWitt equation defined on the full
phase space. We shall apply the reduced approach to these models in
the next section. 

Note that in the case of the tachyon field discussed above, the
demand for a self-adjoint
Hamiltonian does not impose any restrictions on the wave
function which is a solution of (\ref{WDW3}).  

If we demanded the avoidance of both the Big Bang and the Big Brake singularities
in the sense of the DeWitt criterion, we would have to impose the boundary conditions
$\psi(p_T,0)=0$ {\em and} $\psi(p_T,(2\vert p_T\vert/W_0)^{1/3})=0$. The situation would then
be analogous to that of a non-relativistic particle in an infinite potential well,
which is known to lead to a discrete spectrum. In our case, this would lead to discrete
tachyon momenta $p_T=\pm\vert p_T\vert_n$, $n\in {\mathbb N}$. 

For the case of a more general potential than the constant one,
the quantization becomes complicated, for $T$ and $p_T$ appear simultaneously under the square root. 
In the following, we show one possibility how to deal with this problem.
We first perform the canonical transformation
\begin{equation}
	\begin{aligned}	
		T & \rightarrow \phi := \int V(T)\  dT, \\
		p_T & \rightarrow p_{\phi}:= \frac{p_T}{V(T)}.
	\end{aligned}
	\label{canonical1}	
\end{equation} The Hamiltonian constraint then takes the form
\begin{equation} 
	\mathcal{H}=-\frac{p_a^2}{2a}+\sqrt{p_{\phi}^2+a^6}V(T(\phi)). 
\end{equation}
Thus the transformation enables us to move $V$ out of the square root. In the following,
we specialize to potentials of the form
$V(T)=V_1 T^n$, where $n \neq -1$. According to (\ref{canonical1}),
we obtain $\phi=\frac{V_1}{n+1}T^{n+1}$, and therefore the Hamiltonian constraint becomes
\begin{equation} 
\mathcal{H}=-\frac{p_a^2}{2a}+V_1
\left(\frac{(n + 1)\phi}{V_1}\right)^{\frac{n}{n + 1}}
\sqrt{p_{\phi}^2+a^6}.
\end{equation}
After quantization and imposing a simple factor ordering, we get 
\begin{equation} 
\left[
\dif{}{a}{2}
+\mu a\sqrt{p_{\phi}^2+a^6} \left(\dif{}{p_\phi}{}\right)^{\frac{n}{n + 1}}
\right]\Psi (p_\phi,a)=0,
\end{equation}
where $\mu := 2V_1\left(\frac{\I(n + 1)}{V_1}\right)^{\frac{n}{n + 1}}$. 
The Wheeler-DeWitt equation is thus a fractional partial differential equation, 
which can be well defined in the sense of fractional calculus; see, for example, \cite{fractional}.
Note, however, that fractional derivatives can be represented as integral (and thus non-local) operators. 

For the special case of the inverse square potential ($n=-2$), the Wheeler-DeWitt equation becomes
\begin{equation}
\left[
\dif{}{a}{2}
-\frac{2a}{V_1}\sqrt{p_{\phi}^2+a^6} \dif{}{p_\phi}{2}
\right]\Psi (p_\phi,a)=0.
\end{equation}
This is a wave equation with variable coefficients. 
In the region where $a^6 \ll p_\phi^2$, it assumes the asymptotic form
\begin{equation}	\left[
	\dif{}{a}{2}
	-\frac{2a|p_\phi|}{V_1} \dif{}{p_\phi}{2}
	\right]\Psi (a,p_\phi)=0.
	\end{equation}
The separation ansatz $\Psi (a,p_\phi)=\chi(a) \varphi (p_\phi)$ yields the solutions

\begin{equation}
\begin{aligned}
\chi (a) = &
b_1 \Ai \left( \left[\frac{2\lambda}{V_1}^\frac{1}{3}a \right]\right)+
b_2 \Bi \left( \left[\frac{2\lambda}{V_1}^\frac{1}{3}a \right]\right), \\
\varphi(p_\phi)= &
c_1 \sqrt{p_\phi}\; I_1 \left( 2 \sqrt{\lambda |p_\phi |}\right)+
c_2 \sqrt{p_\phi}\;K_1 \left( 2 \sqrt{\lambda |p_\phi |}\right),
\end{aligned}
\end{equation}
where $b_1,b_2,c_1,c_2 \in \mathbb{C}$, and $\lambda\in \mathbb{C}$ is a separation constant; $I_1$ and $K_1$ are the modified Bessel functions.
In the region where $a^6 \gg p_\phi^2$, the asymptotic form of the Wheeler-DeWitt equation is
\begin{equation}	\left[
	\dif{}{a}{2}
	-\frac{2a^4}{V_1} \dif{}{p_\phi}{2}
	\right]\Psi (a,p_\phi)=0.
\end{equation}
A separation ansatz of the form $\Psi (a,p_\phi)=\tilde{\chi}(a) \tilde{\varphi} (p_\phi)$ then yields
\begin{equation}
\begin{aligned}
\bar{\chi} (a) = &
d_1 \sqrt{a}J_{\frac{1}{6}}\left(\sqrt{\frac{2\tilde{\lambda}}{V_1}}\frac{a^3}{3}\right)+
d_2 \sqrt{a}J_{-\frac{1}{6}}\left(\sqrt{\frac{2\tilde{\lambda}}{V_1}}\frac{a^3}{3}\right), \\
\bar{\varphi}(p_\phi)= &
f_1 \exp \left(\sqrt{\tilde{\lambda}}p_\phi\right)+
f_2 \exp \left(-\sqrt{\tilde{\lambda}}p_\phi\right),
\end{aligned}
\end{equation}
where $d_1,d_2,f_1,f_2,\tilde{\lambda}  \in \mathbb{C}$.

If applied to the case of the pseudotachyon, the above procedure leads to the Wheeler-DeWitt equation 
\begin{equation}
\left[
\dif{}{a}{2}
-\frac{2a}{W_1}\sqrt{p_{\phi}^2-a^6} \dif{}{p_\phi}{2}
\right]\Psi (p_\phi,a)=0.
\end{equation}
In the region where $a^6 \ll p_\phi^2$, the asymptotic solutions are the same as in the tachyon case. 
An open question is the behavior of the wave function near the Big Brake singularity.
Using the same method as in the constant $W$ case does not work here, 
since $p_{\phi}$ cannot be treated as a fixed parameter anymore.

\section{Reduced phase space quantization}

In this section, we shall study the cosmological models with the
Born-Infeld type fields with a constant potential, using  
the reduction to physical degrees of freedom approach
\cite{Blyth-Isham,barvin,BarKam2014}. In the case of tachyons, one can
choose as a time parameter $\tau$ 
the cosmological radius $a$ (such a choice is sometimes called an
intrinsic time choice). Indeed, in this case the classical evolution
of the universe is such that the cosmological radius changes
monotonically from the Big Bang to an infinite expansion or from an
infinite contraction ending in the Big Crunch singularity. As a matter
of fact, it is convenient to choose $\tau=a$ for the expansion
and $\tau = -a$ for the contraction. (We choose here the letter $\tau$
to avoid confusion with the classical time parameter $t$.) 

Let us first choose the case 
\begin{equation}
\tau = a.
\label{time}
\end{equation}
Then, the effective non-vanishing  Hamiltonian in the reduced phase
space of the physical degrees of freedom is given by the
corresponding conjugate momentum $p_a$, taken with the inverse sign and
expressed in terms of the physical degrees of freedom and $\tau$,
\begin{equation}
{\cal H}_{\mathrm{red}} = -p_a=+\sqrt{2\tau\sqrt{p_T^2+\tau^6V_0^6}}. 
\label{time1}
\end{equation}
The chosen negative sign of the momentum $p_a$ corresponds to the
expansion of the universe, cf. \eqref{momenta}.
 The expression for this Hamiltonian is well
defined for $0\leq \tau < \infty$. The general solution of the
Schr\"odinger equation corresponding to \eqref{time1} is  
\begin{equation}
\psi(p_T,\tau)=\psi(p_T,0)
\exp\left(-\I\int_0^{\tau}\D\tilde{\tau}\ \sqrt{2\tilde{\tau}\sqrt{p_T^2+\tilde{\tau}^6V_0^6}}\right). 
\label{time2}
\end{equation}
Here, we have used the fact that the time-dependent Hamiltonian
(\ref{time1}) commutes with itself at different moments of
time. Otherwise, it is necessary to use the chronological
T-exponentiation, which makes the formalism more involved.    

Analogously, to describe a contracting quantum universe, it is
convenient to choose the time parameter as 
\begin{equation}
\tau = -a.
\label{time3}
\end{equation}
Then the Hamiltonian is 
\begin{equation}
{\cal H}_{\mathrm{red}} = -p_a=-\sqrt{-2\tau\sqrt{p_T^2+\tau^6V_0^6}},
\label{time4}
\end{equation}
and it is well-defined for $-\infty < \tau \leq 0$. The 
solution of the Schr\"odinger equation is
\bea
& & \psi(p_T,\tau)=\psi(p_T,-\infty)\times\nonumber\\ \ \ \  & & \exp\left(+\I\int_0^{\tau}
\D\tilde{\tau}\ \sqrt{-2\tilde{\tau}\sqrt{p_T^2+\tilde{\tau}^6V_0^6}}\right).
\label{time5}
\eea
The only requirement which one should impose on these solutions is the
normalizability of the wave functions $\psi(p_T,0)$ and
$\psi(p_T,-\infty)$, prescribing  
the initial conditions. The time-dependent part of the solutions
(\ref{time2}) and (\ref{time5}) is simply a phase factor, which
behaves well at the values of time corresponding to  
both the singularities and the infinite volume of the universe.  

For constant potential, the classical momentum
conjugated to the tachyon field is constant, because then the Hamiltonian
(\ref{H-constraint}) is T-independent. Thus, it seems that it is
impossible to describe the dynamics in terms of this momentum, which is the only
observable on which the wave functions (\ref{time2}) and 
(\ref{time5}) depend. However, in quantum theory there is no strong causal
relation between the geometric characteristics of the universe and the
quantities that characterize the matter content.
In our case, one can say that the cosmological radius is not a
geometric characteristic, but a time parameter. Hence, the
cosmological singularity can be associated with a state of the
system corresponding to infinite energy density.
In our case, this means that the time derivative of the
tachyon field tends to  
one, see (\ref{dens}), and the conjugate momentum tends to infinity. It is clear that
the requirement of the  
normalizability of the wave function of the  universe implies the
rapid vanishing of this function at $|p_T| \rightarrow \infty$ and
this fact could be interpreted as a suppression of the  
Big Bang - Big Crunch singularity \cite{KM12}.  

The case of the cosmological model with a pseudotachyon field is more
complicated, because the universe begins its evolution from the Big
Bang singularity, then expands until the occurrence of the Big Brake
singularity, after which it contracts to the
Big Crunch singularity. In such a situation, it is preferable to
employ an extrinsic instead of an intrinsic time \cite{OUP,Kuchar};
an extrinsic time is one that depends on the extrinsic curvature,
see \cite{BarKam2014} and the references
therein. In fact, we cannot use the cosmological radius or a function
of it as a time parameter, because it changes non-monotonically during
the evolution.
  
It is convenient to perform a canonical transformation, which leads
to the new coordinate  
\begin{equation}
q := \frac{p_a}{a^2}. 
\label{time-ex}
\end{equation}
It is easy to check that $q$ is equal to the Hubble
parameter $H={\dot{a}}/{a}$, taken with the opposite sign. If we
identify this new coordinate with the extrinsic time parameter, $\tau\equiv q$, 
the latter is defined in the interval $-\infty < \tau <
+\infty$. The conjugate momentum to $q$ is  
\begin{equation}
p_q = -\frac{a^3}{3}.
\label{time-ex2}
\end{equation}
The reduced Hamiltonian depends on the physical degree of freedom 
$p_T$ and on $\tau$ and is given by
\begin{equation}
{\cal H}_{\mathrm{red}} = -p_q=\frac{a^3}{3} =
\frac{|p_T|}{3\sqrt{\frac14\tau^4+W_0^2}}. 
\label{time-ex3}
\end{equation}
Here, we have used the Hamiltonian constraint to express the
momentum $p_q$  in terms of the physical variable $p_T$ and 
$\tau$. Note that this constraint 
represents a simple quadratic equation with respect to $a^3$, but 
because of the non-negativity of the cosmological radius $a$
we should take the positive square root.
This means that, in contrast to the preceding case of the
tachyon field, we have only one possibility for the choice of the
Hamiltonian and, hence, there exists only one branch of the  
physical wave function of the universe,
\begin{equation}
\psi(p_T,\tau) = \psi(p_T,-\infty)\exp\left(-\I\int_{-\infty}^{\tau}\D\tau\
  \frac{|p_T|}{3\sqrt{\frac14{\tau}^4+W_0^2}}\right).  
\label{time-ex4}
\end{equation}
The existence of only one branch for the wave function
in the reduced approach is in agreement with the fact that the
wave function satisfying the Wheeler-DeWitt
equation, as described in the preceding section, should obey a boundary
condition that guarantees the self-adjointness of the
super-Hamiltonian. Note that the form of the Hamiltonian
(\ref{time-ex3}) is automatically self-adjoint as it should be. The
presence of two branches in the Wheeler-DeWitt wave function or of two
different wave functions, with two different Hamiltonians in the case
of the tachyon field model, is connected with the fact that there are
two different classical cosmological evolutions: expansion and
contraction. 
In the model with a pseudotachyon field we have only one Hamiltonian
for the Schr\"odinger     
equation in the reduced space and an additional boundary
condition for the solution of the Wheeler-DeWitt equation. This
corresponds to only one type of cosmological evolution in this model --
from the Big Bang to the Big Crunch, passing through the point of
maximal expansion where the universe crosses the Big Brake
singularity.    

Speaking about singularities, we can say that the same arguments
which we have used in  
analyzing the relation between the wave function and the
Big Bang singularity in the model with the tachyon field
can be applied to the case of the pseudotachyon field as well. The situation
with the Big Brake singularity is different. Its appearance is not
connected with some particular behavior of the momentum $p_T$ and it
is not suppressed by the wave function of the universe. This seems
natural, because classically a universe can pass through this
singularity without any difficulty,
see e.g. \cite{KGKGP,Kam-CQG} and the references therein. 

We finally note that the reduced approach becomes rather complicated
in the general situation of a non-constant potential.

\section{Quantum cosmology and difference equations}\label{quantum} 

In this section, we shall perform a canonical transformation in
such a way that the square root in the Hamiltonian disappears,
making the problem more tractable. Here, then, we can
use the analogy that was already mentioned in the
introduction; this analogy concerns 
the quantum mechanics of a collapsing (expanding) thin shell
\cite{BKKT88,Berezin97,Hajicek92,BNT05}. Identifying the
Hamiltonian of the system under consideration with the physical mass,
one obtains there an expression for the Hamiltonian which contains a
hyperbolic function of the momentum operator. Because the exponent of
the momentum operator is the generator of spatial translation, one
then arrives at finite difference equations for the wave function. 

In cosmology, we can try to follow this analogy and perform a transition to
new canonical variables and momenta such that the new Hamiltonian
will be free of square roots and will instead contain a combination of
translation operators. Hence, the Wheeler-DeWitt equation will become a
finite difference equation; more precisely, a mixed
difference-differential equation. Such difference equations are common in
loop quantum cosmology \cite{Bojowald,bianchi}, but have so far not been
discussed in the framework of standard quantum cosmology.    

To be concrete, we introduce a new canonical momentum ${\cal P}$ by 
\begin{equation}
p_T =: a^3V(T)\sinh{\cal P}.
\label{mom-new}
\end{equation}
The reason for this choice
is that it turns the square root in \eqref{H-constraint} into the expression
$a^3V\cosh{\cal P}$. Such a form for the kinetic term has been found 
in the above-mentioned papers \cite{BKKT88,Berezin97,Hajicek92,BNT05}.

We now have to construct the corresponding new canonical coordinate ${\cal
  Q}$ such that  
\begin{equation}
 \{{\cal Q}, {\cal P}\} = 1.
 \label{Pois}
\end{equation}
Generally, the tachyon field $T$ can depend on 
${\cal Q}$ as well as on ${\cal P}$.  
Then,
\begin{eqnarray}
& &\{T,p_T\} = \frac{\partial T}{\partial {\cal Q}}\times a^3V(T)\cosh{\cal P}\nonumber\\
& & +\frac{\partial T}{\partial {\cal Q}}\times a^3
\frac{dV(T)}{dT}\frac{\partial T}{\partial {\cal P}}\sinh{\cal P}\nonumber\\
& & -\frac{\partial T}{\partial {\cal P}}\times a^3\frac{dV(T)}{dT}
\frac{\partial T}{\partial {\cal Q}}\sinh{\cal P} = 1,
\label{mpm-new1}
\end{eqnarray}
which gives the condition
\begin{equation}
 a^3V(T)\cosh{\cal P}\frac{\partial T}{\partial {\cal Q}}=1.
 \label{can-new}
\end{equation}
The last equation can be rewritten as 
\begin{equation}
 V(T)\D T = \frac{d{\cal Q}}{a^3 \cosh{\cal P}}.
 \label{master}
\end{equation}
The canonical coordinate $\mathcal{Q}$ can thus be written as
\begin{equation}
{\cal Q}=a^3 \cosh{\cal P}\int V(T) \D T.
\label{master1_2}
\end{equation}
What about the modification of $a$ and $p_a$? It is convenient to keep the scale factor
as the configuration variable. Unfortunately, this is not possible for its momentum.
The reason is that the Poisson brackets between $p_a$ and $T$ and
between $p_a$ and $p_T$ must vanish. This is not the case for an un-modified $p_a$.
One can easily see that the transformation
\be
\lb{tildepa}
p_a \to \tilde{p}_a:=p_a-\frac{3{\cal Q}\tanh{\cal P}}{a}
\ee
leads to the vanishing of those brackets; that is, 
the new variables $a,{\cal Q},\tilde{p}_a,{\cal P}$
arise from the old ones $a,T,p_a,p_T$ by a canonical transformation. 

In the terminology of \cite{Goldstein}, Sec.~9.1, the generator of this canonical
transformation reads
\be
F_2(a,\tilde{p}_a;T,{\cal P})=a^3\sinh{\cal P}\int V(T)\D T+a\tilde{p}_a,
\ee
where
\be
p_T=\frac{\partial F_2}{\partial T},\quad p_a=\frac{\partial F_2}{\partial a},
\ee
and
\be
{\cal Q}=\frac{\partial F_2}{\partial{\cal P}},\quad a=\frac{\partial F_2}{\partial \tilde{p}_a}.
\ee
In principle, with the help of this generating function,
one may use the method discussed in \cite{Ghandour} to relate 
the corresponding wave functions.

After the canonical transformation, the Hamiltonian constraint \eqref{H-constraint}
assumes the form
\be
\lb{H-constraint2}
{\mathcal H}\equiv -\frac{1}{2a}\left(\tilde{p}_a+\frac{3{\cal Q}\tanh{\cal P}}{a}
\right)^2+a^3V\cosh{\cal P}=0.
\ee
In a particular factor ordering, the
Wheeler-DeWitt equation can now be written as 
\bea
& & 
-\frac{1}{2a}\left(\frac{\hbar}{\I}\frac{\partial}{\partial a}
+\frac{3{\cal Q}}{a}\tanh\left(\frac{\hbar}{\I}\frac{\partial}{\partial {\cal Q}}\right)
\right)^2\psi(a,{\cal Q})\nonumber\\
& & +a^3V\cosh\left(\frac{\hbar}{\I}\frac{\partial}{\partial {\cal Q}}\right)
\psi(a,{\cal Q})=0,
 \label{WDW}
\eea
where $V=V(T(a,{\cal Q},{\cal P}))$.  
This equation has a rather complicated form. We note that
$\tanh\left(\frac{\hbar}{\I}\frac{\partial}{\partial {\cal Q}}\right)$ is not a suitable operator,
since the series expansion of $\tanh$ has a finite radius of convergence. 
But since $\mathcal{H}$ is a constraint, we can multiply it by an arbitrary factor
and the resulting quantity will be a constraint as well.
If we define $ \tilde{\mathcal{H}}:=2a^3\cosh^2(\mathcal{P})\mathcal{H}$ and
perform the transformation $a\rightarrow\alpha=\ln a$, $\tilde{p}_a\rightarrow \tilde{p}_\alpha=a \tilde{p}_a$, 
we obtain the new constraint
\begin{align}
	\tilde{\mathcal{H}}  := & -\tilde{p}_\alpha ^2 \cosh ^2 \mathcal{P} + 
	9{\cal Q}^2 \sinh ^2 \mathcal{P}  +6 \mathcal{Q}\tilde{p}_{\alpha} \sinh \mathcal{P}
	 \cosh 	\mathcal{P} \nonumber \\
	 & +2 e^{6 \alpha}V \cosh ^3 \mathcal{P} .	
\end{align}
We emphasize that the hyperbolic functions with the momentum as argument generate
a translation in the argument; we have, for example,
\bea
& & \cosh\left(-\I\frac{\partial}{\partial {\cal Q}}\right)\psi(\tilde{p}_{\alpha},{\cal Q})
=\frac{1}{2}\left(e^{-\I\frac{\partial}{\partial {\cal Q}}}+
e^{\I\frac{\partial}{\partial {\cal Q}}}\right)\nonumber\\
& & \ =\frac{1}{2}\left(\psi(\tilde{p}_{\alpha},{\cal Q}-\I)+\psi(\tilde{p}_{\alpha},{\cal Q}+\I)\right)
\eea
After naive factor ordering, setting $\hbar=1$, and returning to the case of constant potential,
the Wheeler-DeWitt equation assumes the following form:
\begin{align}
	\frac{V_0e^{6\alpha}}{4}
	&\psi(\alpha,\mathcal{Q}+3 \I ) \nonumber
	\\
	-\frac{1}{4}
	\left(\tilde{p}_\alpha -3 \mathcal{Q}\right)^2
	&\psi(\alpha,\mathcal{Q}+2 \I ) \nonumber
	\\
	+\frac{3V_0e^{6\alpha}}{4}
	&\psi(\alpha,\mathcal{Q}+\I ) \nonumber
	\\
	+\frac{-\tilde{p}_\alpha ^2 + 9\mathcal{Q} ^2}{2}
	&\psi(\alpha,\mathcal{Q}) \nonumber
	\\
	+\frac{3V_0e^{6\alpha}}{4}
	&\psi(\alpha,\mathcal{Q}-\I ) \nonumber
	\\
	-\frac{1}{4}
	\left(\tilde{p}_\alpha +3 \mathcal{Q}\right)^2
	&\psi(\alpha,\mathcal{Q}-2 \I ) \nonumber
	\\
	+\frac{V_0e^{6\alpha}}{4}
	&\psi(\alpha,\mathcal{Q}-3 \I ) \nonumber
	\\
	=0.
\end{align}
This is a mixed difference-differential equation (or partial difference equation, if we use the momentum representation for 
$\alpha$). 

In the asymptotic limit of large $a$, \eqref{WDW} reads
\be
\frac{\partial^2\psi}{\partial a^2}
 +2a^4V_0\cosh\left(\frac{\hbar}{\I}\frac{\partial}{\partial {\cal Q}}\right)
\psi=0.
\ee
Apart from the cosh-term, this coincides with the earlier form (\ref{WDW5}), which guarantees
the consistency of the formalism.
Employing the product ansatz
\be
\psi(a,{\cal Q})=\phi({\cal Q})\chi(a),
\ee
we find
\bea
\frac{d^2\chi}{da^2} &=& \frac{12}{\kappa^2}a^4V_0\lambda\chi(a),\lb{chi} \\
\cosh\left(-\I\hbar\frac{d}{d{\cal Q}}\right)\phi({\cal Q}) &=&-\lambda\phi({\cal Q}),\lb{phi}
\eea
where we take $\lambda$ to be a real constant. Introducing for convenience 
$\Lambda:=2V_0\lambda$ (recall $V_0>0$), we find for the solutions of
\eqref{chi} a combination of  
Bessel functions. For $\Lambda>0$, we find the solutions
$\sqrt{a}I_{1/6}(\sqrt{\Lambda} 
a^3/3)$ and $\sqrt{a}K_{1/6}(\sqrt{\Lambda}a^3/3)$, while for
$\Lambda<0$, we find 
$\sqrt{a}J_{1/6}(\sqrt{-\Lambda}a^3/3)$ and
$\sqrt{a}J_{-1/6}(\sqrt{-\Lambda}a^3/3)$. We note that for $\lambda=1$ this 
corresponds to the solutions (\ref{Bessel}).

In order to make a selection amongst these Bessel functions, we inspect
their asymptotic behavior. Let us first consider the case $\Lambda >0$. The solution $\sqrt{a}I_{1/6}(\sqrt{\Lambda}
a^3/3)$ increases exponentially with large $a$ and is thus not
normalizable; it must 
be excluded. The solution $\sqrt{a}K_{1/6}(\sqrt{\Lambda}a^3/3)$ decreases
exponentially and is thus normalizable. 
The solutions for $\Lambda<0$, on the other hand, are oscillatory and thus both
allowed; they correspond to (\ref{Bessel}) with $\lambda=-1$. 

The second equation \eqref{phi} can be re-written in the form of the 
following difference equation:
\be
\lb{difference}
\frac12\left[\phi({\cal Q}+\I)+\phi({\cal Q}-\I)\right]=-\lambda\phi({\cal Q}).
\ee
Making the ansatz
\be
\lb{phi-ansatz}
\phi({\cal Q})=e^{\alpha {\cal Q}},
\ee
one finds
\be
\I\alpha_{1,2}=\ln\left(-\lambda\pm\sqrt{\lambda^2-1}\right).
\ee
Inserting this into \eqref{phi-ansatz} and writing
$-\lambda=:\cosh{\cal P}_0$, one gets for $\lambda<0$,
\be
\phi({\cal Q})=d_1e^{-\I{\cal P}_0{\cal Q}}+d_2e^{\I{\cal P}_0{\cal Q}},
\ee
with constants $d_1$ and $d_2$. (This is also expected from the 
momentum representation of the Wheeler-DeWitt equation.) Taking all
this together, 
the most general allowed asymptotic solution for $\Lambda<0$ is given by
\begin{eqnarray*}
& & \psi(a,{\cal Q}) =\left(d_1e^{-\I{\cal P}_0{\cal Q}}+d_2e^{\I{\cal P}_0{\cal Q}}\right)
\times \nonumber\\ 
& & \!\!\!\!\!\!\left[c_1\sqrt{a}J_{1/6}\left(\sqrt{-\Lambda}a^3/3\right)+c_2
\sqrt{a}J_{-1/6}\left(\sqrt{-\Lambda}a^3/3\right)\right]. 
\end{eqnarray*}
For $\Lambda >0$, one obtains
\begin{eqnarray*}
& & \psi(a,{\cal Q}) =\left(e_1e^{-\I{\cal P}_0{\cal Q}}+e_2e^{\I{\cal P}_0{\cal Q}}\right)e^{-\pi {\cal Q}}
\times \nonumber\\ 
& & \!\!\!\!\!\!
\sqrt{a}K_{1/6}\left(\sqrt{\Lambda}a^3/3\right). 
\end{eqnarray*}

Can we say something about the general equation \eqref{WDW}?
In the limit of small $a$, this equation assumes the form
\be
(a\tilde{p}_a)\cosh{\cal P}\psi(a,{\cal Q})+3{\cal Q}\sinh{\cal P}\psi(a,{\cal Q})=0.
\ee
This leads to the difference equation
\bea
& & 
(a\tilde{p}_a)\left[\psi(a,{\cal Q}+\I)+\psi(a,{\cal Q}-\I)\right]\nonumber\\
& & \ -3{\cal Q}\left[\psi(a,{\cal Q}+\I)-\psi(a,{\cal Q}-\I)\right]=0.
\eea
After switching to the variable $\alpha$ and going to momentum space, it reads
\begin{equation*}
\left(\tilde{p}_\alpha -3\mathcal{Q} \right)\psi\left(\tilde{p}_\alpha, 
\mathcal{Q}+i \right)+\left(\tilde{p}_\alpha +3\mathcal{Q} \right)\psi\left(\tilde{p}_\alpha, \mathcal{Q}-i \right)=0.
\end{equation*}
A particular set of solutions is 
\begin{equation}
\label{Gamma}
	\psi\left(\tilde{p}_\alpha, \mathcal{Q} \right)
	= \mu(\tilde{p}_\alpha ,\mathcal{Q})
	\frac{\Gamma\left(
	 -\frac{\text{i}}{2}\left[
	 \frac{\tilde{p}_\alpha}{3}+
	 \mathcal{Q}+\text{i}\right]\right)}
	 {\Gamma\left(
	 \frac{\text{i}}{2}\left[
	 \frac{\tilde{p}_\alpha}{3}-
	 \mathcal{Q}-\text{i}\right]\right)},
\end{equation}
where $\mu(\tilde{p}_\alpha,\mathcal{Q})$ is a function that is 
an arbitrary $2 \text{i}$-periodic function in the second argument, 
that is, $\mu(\tilde{p}_\alpha,\mathcal{Q})=\mu(\tilde{p}_\alpha,\mathcal{Q}+2 \text{i})$. 

We emphasize that the occurrence of the Gamma function in (\ref{Gamma}) is not an accident.
The Gamma function obeys the perhaps most famous difference equation, 
$\Gamma(x+1)=x\Gamma(x)$, and it is known that it does not satisfy any algebraic differential
equation whose coefficients are rational functions; the latter property is known as
H\"older's theorem (see e.g. \cite{BK78,Norlund}). Thus, as emphasized in \cite{Norlund}, one gets
from difference equations transcendental functions of a very different kind than from 
differential equations. 

Let us now consider the case of the inverse $T$-squared potential,
\begin{equation}
 V=\frac{V_1}{T^2},
 \label{two}
\end{equation}
whose classical behavior was discussed in Sec.~II.
For this potential, one can perform the canonical transformation
\begin{equation}
	p_T=a^3V_1\sinh \mathcal{P},
\end{equation}
\begin{equation}
	T=-\frac{a^3 V_1 \cosh \mathcal{P}}{\mathcal{Q}},
\end{equation}
\begin{equation}
	p_a=\tilde{p}_a+\frac{3\mathcal{Q}\tanh \mathcal{P}}{a}.
\end{equation}
The Hamiltonian constraint becomes
\begin{equation}
	\mathcal{H}=	-\frac{1}{2a}\left(\tilde{p}_a + \frac{3\mathcal{Q}\tanh 
	\mathcal{P}}{a}\right)^2
	+\frac{\mathcal{Q}^2}{a^3 V_1 \cosh \mathcal{P}}.
\end{equation}
After another canonical transformation
$a \rightarrow \alpha=\ln a$, $\tilde{p}_a\rightarrow\tilde{p}_\alpha=a \tilde{p}_a$, the Hamiltonian assumes the form
\begin{equation}
\label{newH}
	\mathcal{H} =e^{-3 \alpha}	\left[-\frac{1}{2}\left(\tilde{p}_{\alpha}^2 
	+3\mathcal{Q}\tanh 
	\mathcal{P}\right)^2+
	\frac{{\cal Q}^2}{V_1 \cosh \mathcal{P}}\right].
\end{equation}
We observe that $\frac{\partial\mathcal{H}}{\partial\alpha}=-3\mathcal{H} = 0$.
Thus the canonical momentum $\tilde{p}_{\alpha}=p_{\alpha}- 3 \mathcal{Q}\tanh \mathcal{P}$ is a constant of motion. 
If we reinsert the old coordinates, this leads to
\begin{equation}
	a p_a + 3 p_T T = \text{constant}=: C.
	\label{const_of_motion}
\end{equation}

As a side remark, we want to mention that the canonical transformation 
for the pseudotachyon case looks similar. One just has to replace
$\sinh\rightarrow \cosh$, $\cosh\rightarrow \sinh$ and $\tanh\rightarrow \coth$. 
The constant of motion is then also present (with the same value)
in the pseudotachyon model with inverse $T$-squared potential.

If we employ in the tachyon case the same procedure as for the constant $V$ model, 
we obtain
\begin{align}
\tilde{\mathcal{H}}=& -\tilde{p}_\alpha^2 \cosh ^2 \mathcal{P}- 6{\cal Q} \tilde{p}_{\alpha}\sinh \mathcal{P}
\cosh \mathcal{P}-9{\cal Q}^2 \sinh ^2 \mathcal{P}\nonumber \\ +& \frac{2\mathcal{Q}^2}{V_1} \cosh \mathcal{P}.
\end{align}
With naive factor ordering, the Wheeler-DeWitt equation reads
\begin{align}
	-\frac{1}{4}
	\left(\tilde{p}_\alpha -3 \mathcal{Q}\right)^2
	&\psi(\tilde{p}_\alpha,\mathcal{Q}+2 \I) \nonumber
	\\
	+\frac{\mathcal{Q}^2}{V_1}
	&\psi(\tilde{p}_\alpha,\mathcal{Q}+\I) \nonumber
	\\
	+\frac{-\tilde{p}_\alpha ^2 + 9\mathcal{Q} ^2}{2}
	&\psi(\tilde{p}_\alpha,\mathcal{Q}) \nonumber
	\\
	+\frac{\mathcal{Q}^2}{V_1}
	&\psi(\tilde{p}_\alpha,\mathcal{Q}-\I) \nonumber
	\\
	-\frac{1}{4}
	\left(\tilde{p}_\alpha +3 \mathcal{Q}\right)^2
	&\psi(\tilde{p}_\alpha,\mathcal{Q}-2 \I) \nonumber
	\\
	=0
\end{align}
This is an analytic difference equation. The foundations for such equations are presented in
the book by N\o rlund \cite{Norlund}. 
In order to achieve conformity with N\o rlund's notation, we define
$z:=-\I\mathcal{Q}$ and $u(z):=\psi (\mathcal{Q}(z))$. The function $u$ then fulfills the difference equation
\begin{equation}
	\sum\limits_{k=0}^4 P_k(z) u(z+k)=0,
	\label{diff_eq2}
\end{equation}
where the polynomials $P_k$ are given by 
\begin{equation*}
\begin{aligned}
P_0(z)=\frac{1}{4}(&\tilde{p}_\alpha -3 \I z)^2, \; 
P_1(z)=\frac{z^2}{V_1}, \; 
P_2(z)=\frac{\tilde{p}_\alpha ^2 +9z^2}{2}, \\
&P_3(z)=\frac{z^2}{V_1}, \;
P_4(z)=\frac{1}{4}(\tilde{p}_\alpha +3 \I z)^2.
\end{aligned}
\end{equation*}
The problem of solving the difference equation (\ref{diff_eq2}) is equivalent to the problem of 
finding a fundamental system of solutions. The solution space is a subset of the meromorphic functions 
and the fundamental system will be a vector space over the field of $1$-periodic meromorphic  functions.
Following N\o rlund, we expect the fundamental system to be composed of four linearly independent functions.
Note that $\deg P_k=2$ for all $k=0,1,\dots,4$, 
and thus the difference equation fulfills the first criterion to belong to a certain class 
which N\o rlund calls normal difference equations. To check the second criterion we have to rewrite
\begin{equation}
P_k(z)=\sum\limits_{l=0}^2 c_{k,l}\prod\limits _{n=0}^{l-1} (z+k+n)
\end{equation}
The second criterion demands the non-degeneracy of the zeros $a_n$ of the polynomial $f_2(s)$, where
\begin{equation}
f_l(s):=\sum_{k=0}^4 c_{k,l} s^k
\end{equation}
The zeros are given by complicated expression. However, they are non-degenerate.
N\o rlund shows explicitly that the solutions to normal difference equations 
can be written in terms of products of Gamma functions and a series of rational functions
(Chap.~11, Sec.~2 in \cite{Norlund}). In our case, the solutions will be of the form
\begin{equation}
u_{n}(z)=a_n^z\frac{\Gamma(z)}{\Gamma(-\beta_{n})\Gamma(z+\beta_{n}+1)}\Omega(z,\beta_{n}),
\end{equation}
where
\begin{equation}
\Omega(z,\beta_{n})=\sum\limits_{\nu =0}^{\infty}A_{\nu}\frac{(\beta_{n}+1)(\beta_{n}+2)\dots (\beta_{n}+\nu)}{(z+\beta_{n}+1)\dots(z+\beta_{n}+\nu)}
\end{equation}
and $n=1,2,3,4$.
The coefficients $A_{\nu}$ can in principle be determined by plugging the solutions into (\ref{diff_eq2}). 
The coefficients  $\beta_n$, however, depend on the explicit form of 
the singular solutions of a certain differential equation in a neighborhood of the $a_n$'s
and cannot be determined easily. Nevertheless, the above discussion shows how difference equations
occuring in quantum cosmology can in principle be dealt with.

We finally mention that, with the Hamiltonian (\ref{newH}),
wet get in the momentum representation an ordinary differential equation.

\section{Conclusion}

The purpose of our paper is to investigate a certain class of cosmological models
with Born-Infeld (tachyon) type of fields. This is of interest for (at least) two reasons.
First, such models have been encountered in the study of models for Dark Energy.
Second, the Lagrangian is of a square-root type and thus poses challenges for quantization,
which we have discussed here in detail. In the classical part, we have in particular obtained
new results concerning the occurrence of a Big-Brake singularity for the inverse square potential
in the pseudotachyon case.
Concerning quantization, we have managed to get non-trivial results
for two models: for the model with a constant tachyon (pseudotachyon) potential, which is
equivalent to the Chaplygin (anti-Chaplygin) gas and for the model where 
the potential is inversely proportional to the tachyon field squared. 
We have derived and discussed the Wheeler--DeWitt equation for these models.
For constant potential,
it is convenient to use the momentum representation for the quantum state and 
to consider the reduced phase space of physical variables. This becomes quite complicated 
for non-constant potentials. We have thus pointed out a general method to transform the Wheeler--DeWitt equation
into a form without square roots. This leads to a difference equation,
which requires new methods for its solution. We have discussed these methods and pointed out 
that difference equations lead in general to solutions of a different type than solutions
from differential equations. We have outlined a general procedure to finding such solutions.
The methods may also be of use in loop quantum cosmology
\cite{Bojowald,bianchi}.

\section*{Acknowledgments}
We gratefully acknowledge financial support by the Foundational Questions
Institute (www.fqxi.org). C.K. thanks the Max Planck Institute for
Gravitational Physics (Albert Einstein Institute), Potsdam, Germany,
for kind hospitality while part of this work was done.
The work of A.K. was partially supported by the RFBR through the grant
No 14-02-00894.  



\end{document}